\documentclass{aastex}
\usepackage{natbib}
\newcommand{\hii}{H~{\sc ii}}
\def\farcs{\hbox{$.\!\!^{\prime\prime}$}}
\newcommand{\rstar}{r_\mathrm{star}}
\newcommand{\tstar}{T_\mathrm{star}}
\newcommand{\kb}{k_\mathrm{B}}
\newcommand{\tauion}{\tau_\mathrm{ion}}
\newcommand{\taupm}{\tau_\mathrm{Pm}}
\newcommand{\kp}{\kappa_\mathrm{P}}
\newcommand{\gd}{\Gamma_\mathrm{d}}
\newcommand{\gacc}{\Gamma_\mathrm{acc}}
\newcommand{\gs}{\Gamma_\mathrm{st}}
\newcommand{\racc}{r_\mathrm{acc}}
\newcommand{\tacc}{T_\mathrm{acc}}
\newcommand{\lacc}{L_\mathrm{acc}}
\newcommand{\rs}{R_\mathrm{S}}
\newcommand{\cs}{c_\mathrm{s}}
\newcommand{\tmin}{T_\mathrm{min}}
\newcommand{\mpr}{m_\mathrm{p}}
\newcommand{\nel}{n_\mathrm{e}}

\begin{document}

\title{\hii\ regions: Witnesses to massive star formation}

\author{Thomas Peters, Robi Banerjee and Ralf S. Klessen}
\affil{Zentrum f\"{u}r Astronomie der Universit\"{a}t Heidelberg,  
Institut f\"{u}r Theoretische Astrophysik, Albert-Ueberle-Str. 2, D-69120 Heidelberg, Germany}
\email{thomas.peters@ita.uni-heidelberg.de}

\author{Mordecai-Mark Mac Low}
\affil{Department of Astrophysics, American Museum of Natural History,
79th Street at Central Park West, New York, New York 10024-5192, USA}

\author{Roberto Galv{\'a}n-Madrid\altaffilmark{1,2} and Eric R. Keto}\affil{Harvard-Smithsonian Center for Astrophysics, 60 Garden Street, Cambridge, MA 02138, USA}

\altaffiltext{1}{Centro de Radioastronom{\'\i}a y Astrof{\'\i}sica, UNAM, A.P. 3-72 Xangari, Morelia 58089, Mexico}
\altaffiltext{2}{Academia Sinica Institute of Astronomy and Astrophysics, P.O. Box 23-141, Taipei 106, Taiwan}

\begin{abstract}

  We describe the first three-dimensional simulation of the
  gravitational collapse of a massive, rotating molecular cloud that
  includes heating by both non-ionizing and ionizing radiation. These
  models were performed with the FLASH code, incorporating a hybrid,
  long characteristic, ray tracing technique.  We find that as the
  first protostars gain sufficient mass to ionize the accretion flow,
  their \hii\ regions are initially gravitationally trapped, but soon
  begin to rapidly fluctuate between trapped and extended states, in
  agreement with observations.  Over time, the same ultracompact \hii\
  region can expand anisotropically, contract again, and take on any
  of the observed morphological classes. In their extended phases,
  expanding \hii\ regions drive bipolar neutral outflows
  characteristic of high-mass star formation. The total lifetime of
  \hii\ regions is given by the global accretion timescale, rather
  than their short internal sound-crossing time. This explains the
  observed number statistics.  The pressure of the hot, ionized gas
  does not terminate accretion.  Instead the final stellar mass is set
  by fragmentation-induced starvation. Local gravitational
  instabilities in the accretion flow lead to the build-up of a small
  cluster of stars, all with relatively high masses due to heating
  from accretion radiation.  These companions subsequently compete
  with the initial high-mass star for the same common gas reservoir
  and limit its mass growth. This is contrary to the classical competitive
  accretion model, where the massive stars are never hindered in growth
  by the low-mass stars in the cluster. Our findings show that the most
  significant differences between the formation of low-mass and
  high-mass stars are all explained as the result of rapid accretion
  within a dense, gravitationally unstable, ionized flow.

\end{abstract}

\maketitle

\section{Introduction}

Massive stars influence the surrounding universe far out of proportion
to their numbers through ionizing radiation, supernova explosions, and
heavy element production. Their formation requires the collapse of
massive interstellar gas clouds with accretion rates
exceeding $10^{-4}$~M$_{\odot}$~yr$^{-1}$ \citep{beutheretal02,beltranetal06}
to reach their final masses before exhausting their nuclear fuel \citep{ketoetal06}.
Massive stars can ionize the gas around them, forming \hii\ regions, that
traditionally have been modelled as spherical bubbles expanding in a
uniform medium \citep{spitzer78}.  This approach, however, fails to
explain their observed numbers and morphologies
\citep{woodchurch89,kurtzetal94}. A more recent alternative picture
models \hii\ regions as the ionized section of the accretion flow that
feeds the young massive star \citep{keto02,keto07}. The rich diversity
of observed \hii\ regions thus bears witness to the complexity of
high-mass star formation.

\hii\ regions form around accreting protostars once they exceed $\sim
10\,$M$_\odot$, equivalent to a spectral type of early B.  Thus,
accretion and ionization must occur together in the formation of
massive stars. The pressure of the 10$^4$~K ionized gas far exceeds
that in the 10$^2$~K accreting molecular gas, producing feedback in
the form of ionized outflows \citep{keto02,keto03,keto07}.

High-mass stars form in denser and more massive cloud cores
\citep{motteetal08} than their low-mass counter\-parts
\citep{myersetal86}. High densities also
result in local gravitational instabilities in the accretion flow,
resulting in the formation of multiple additional stars
\citep{klesbur00,krattmatz06}.  Young massive stars are almost always
observed to have companions \citep{hohaschik81}, and the number of
their companions significantly exceeds those of low-mass stars
\citep{zinnyork07}.  Such companions influence subsequent accretion
onto the initial star \citep{krumholzetal09}.
Observations show an upper mass limit of about $100\,$M$_{\odot}$. It remains unclear
whether limits on internal stability or termination of accretion by stellar feedback
determines the value of the upper mass limit \citep{zinnyork07}.

Around the most luminous stars, the outward radiation pressure
can counterbalance the inward gravitational attraction. A spherically symmetric
calculation of radiation pressure on dust yields equality at just
under 10 M$_\odot$ \citep{wolfcas87}. However, the dust opacity is
wavelength dependent, the accretion is non-spherical, the
mass-luminosity ratio is different for multiple companions than for a
single star, and  the momentum of the accretion flow must be reversed, rather
than just achieving static force
balance \citep{larsstarr71,kahn74,yorkekruegel77,nakanoetal95,sigalottietal09}. Observations
\citep{chini04,pateletal05,ketoetal06,beltranetal06}
provide evidence for the presence of all these mitigating
factors, and numerical experiments combining some of these
effects \citep{yorke02,krumkleinmckee07} confirm their effectiveness, showing
that radiation pressure is not dynamically significant below the Eddington limit.

The most significant differences between massive star formation and
low-mass star formation seem to be the clustered nature of star
formation in dense accretion flows and the ionization of these flows.
We present the first three-dimensional simulations of the collapse
of a molecular cloud to form a cluster of massive stars
that include ionization feedback,
allowing us to study these effects simultaneously.  
During recent years the problem of massive star formation was
tackled in a number of three-dimensional numerical studies \citep[see e.g.][
for a recent review]{klesseneetal09}. However, none of
these simulations included the self-consistent ionization feedback
from massive protostars that dynamically condense out of
gravitationally unstable regions. \citet{daleetal05} performed the
first hydrodynamic calculation of a star cluster forming region to
include a simplified treatment of ionization feeback, but only from a
single source that was inserted into a pre-existing simulation and
whose ionizing flux did not depend on the mass of the growing
(proto-)star. Also, heating by the accretion luminosity and from the
stellar photosphere is neglected in these calculations, but is
included in ours.
Other calculations that incorporate ionizing
radiation focus on triggered star formation and turbulence by external
sources~\citep{daleetal07a,daleetal07b,gritetal09b}. 
Using a flux-limited diffusion approximation,
\citet{krumkleinmckee07} incorporated radiation feedback in their
calculations of massive star formation. These calculations showed that
radiation pressure from non-ionizing radiation cannot halt accretion
onto the massive protostar and therefore is not able to set an upper
mass limit of such stars \citep[see
also][]{krumholzetal09}. Nevertheless, these simulations lack the
important feedback from ionizing radiation distinctive of massive
star formation.

In the following, we present our numerical method  (\S~\ref{S:numerics}) and give a
detailed analysis of the simulation results. We discuss the accretion
history of the stellar cluster (\S~\ref{accretion}), the fragmentation of the rotationally flattened
structure, and the generation of ionization-driven bipolar outflows
(\S~\ref{bipolar}), and compare
synthetic radio continuum and line observations with observed data (\S~\ref{comparison}).
Finally, we discuss the relation of our numerical model to previous
work (\S~\ref{discussion}), and the relevance of our findings for
massive star formation (\S~\ref{conclusions}).

\section{Numerical Method}
\label{S:numerics}

\subsection{Algorithm}
We use a modified version of the FLASH code
\citep{fryxell00}, an adaptive-mesh code that integrates the
compressible gas dynamic equations with self-gravity and radiation
feedback. Our modifications include the propagation of both ionizing
and non-ionizing radiation to follow heating, though not radiation
pressure, a refinement criterion to resolve the Jeans length
\citep{banerjee04}, various cooling processes of relevance for
protostellar collapse \citep{banerjee04,banerjee06b,banerjee06}; and
Lagrangian sink particles \citep{banerjeeetal09,federrathetal09}.
Here we report on our new modifications to the code.

\subsubsection{Radiation}
To propagate radiation, we use the hybrid characteristics raytracing
module originally developed by \citet{rijk06}. Parallel
raytracing on a block-structured adaptive mesh requires the solution
of an inverse problem: Given the position in the domain, the
corresponding block identifier is needed. This problem was originally
approached by storing the block identifier in one single large
array corresponding to a fully refined domain.  Each entry in this
array could be mapped to a (potential) cell in the domain and vice
versa, so that the block identifier could be obtained by looking up
the stored value. This solution is straightforward and fast, but it
violates the principles of adaptive mesh refinement and becomes
extremely memory consuming at high resolution. Collapse simulations
are impossible using this method.

Instead of creating one large array, we use the fact that FLASH stores
hierarchical information about its block structure. Every block in the
adaptive-mesh hierarchy has information about its parent and child
blocks as well as on its neighbors at the same level of
refinement. Valid data is stored for blocks of highest refinement
only, which are called leaf blocks. Neighboring leaf blocks can
differ only by one level of refinement.  This hierarchical information
makes it possible to perform a tree walk in the adaptive-mesh
hierarchy.

Starting from a block whose identifier is known, we determine the
direction in which the ray leaves the block. If there is a
neighboring block on the same level of refinement, it can either
still be a leaf block or have one level of children. In the former
case the new block is already found, in the latter case the child
blocks must be checked.  If there is no neighboring block at the same
refinement level, the new block must be the neighbor of the parent of
the current block in this direction.

Although the method is more complicated than the original one, it is
equally fast since the hierarchical data can be effectively stored and
communicated among processors.

The radiation module has been tested for its capability of accurately
casting shadows \citep{rijk06} and tracing R-type ionization fronts in a
cosmological setting \citep{iliev06}. For simulations of massive star
formation, the expansion of D-type ionization fronts should be
adequately modeled. We have tested the ionization physics in our model
against an approximate solution found by \citet{spitzer78}
for the expansion of a D-type ionization front into a
homogeneous medium. Given the Str\"{o}mgren radius $\rs$ and the sound
speed of the ionized gas $\cs$, the radius $R$ of the ionization front
at time $t$ is given by
\begin{equation} \label{stromgren}
R(t) = \rs \left(1 + \frac{7}{4} \frac{\cs t}{\rs}\right)^{4/7}  .
\end{equation}

\begin{figure}
\includegraphics[height=340pt]{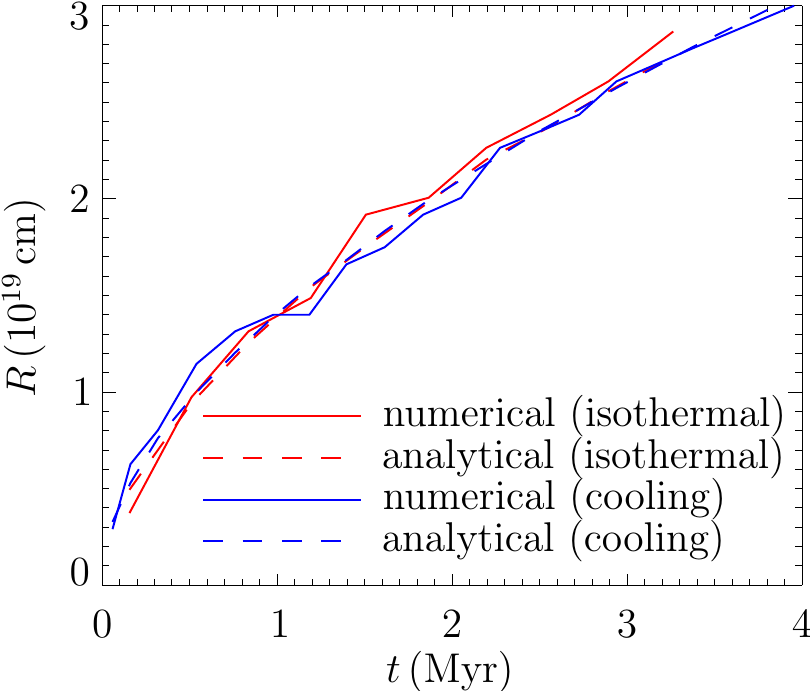}
\caption{Expansion of a D-type ionization front in a homogeneous medium. The plots shows the
radius $R$ of the ionization front as function of time $t$. The analytical data is dashed,
the numerical data is solid. The plot shows a run with an isothermal equation
of state ($\gamma = 1$) without cooling and a run with $\gamma = 5/3$ and a cooling curve.
The results agree well, both with the analytic solution, and with
each other.}
\label{spitzercomp}
\end{figure}

A comparison of our numerical simulation with this analytic solution
is plotted in Fig.~\ref{spitzercomp}. 
We show the data from two different simulations, both of which had initial
density $\rho = 3.3 \times 10^{-21}$~g~cm$^{-3}$, ionizing luminosity
$S_* = 1.4 \times 10^{49}$~s$^{-1}$, and were run with
a maximum resolution of $0.25$~pc. In the first run, we used an isothermal
equation of state ($\gamma = 1$) and included no cooling
processes. In the second, we set $\gamma = 5/3$
and used the cooling curve from \citet{dalgmc72}. In both simulations, the
ionization fraction and temperature of the ionized gas was determined
self-consistently by solving a rate equation for the ionization fraction
as discussed in \citet{rijk06}.

We compare the numerical
result in each case with the result from equation~(\ref{stromgren})
using the appropriate sound speed in the ionized gas $c_s = (\gamma k
T_i / \mu)^{1/2}$, where $k$ is Boltzmann's constant and $\mu$ is the
mean mass per particle.  The agreement of each
run with the analytical solution and with each other
is acceptable. The decent agreement even in the presence of cooling
demonstrates that numerical diffusion 
from the cold, dense, neutral shell into the hot, ionized interior does not
lead to significant unphysical cooling of the hot interior.  

\begin{figure}
\includegraphics[height=300pt]{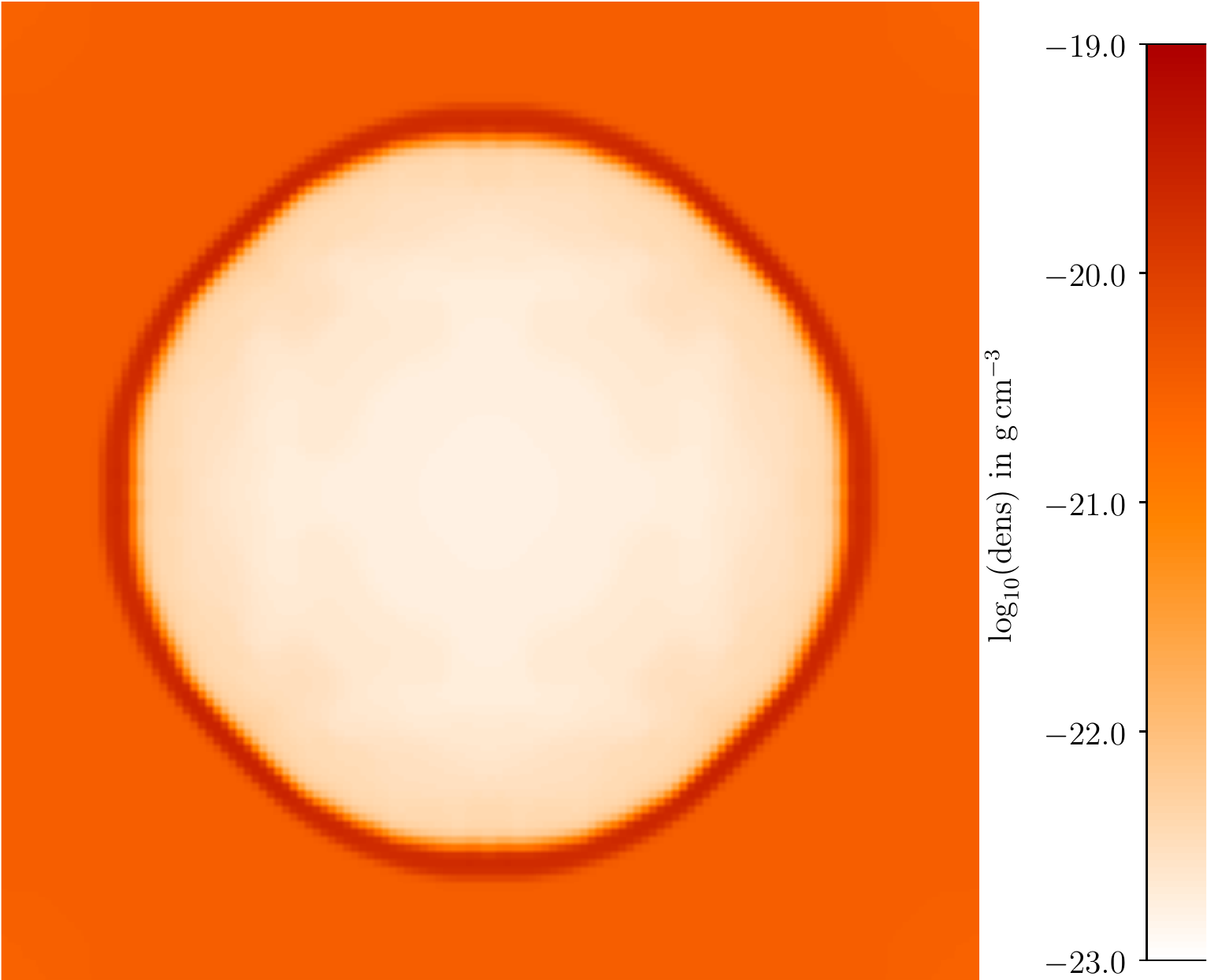}
\caption{Log density slice of an ionization front in a homogeneous medium. The shell 
deviates a little from a perfect sphere due to waves running through the ionized gas,
amplifying grid effects.
The deviations from the spherical shape decrease with increasing resolution.}
\label{spitzerslice}
\end{figure}

We show a slice of the expanding \hii\ region in figure~\ref{spitzerslice}.
The whole box shown has an effective numerical resolution of $128^3$ grid points. Small deviations
from a perfectly spherical shape are visible. This is because the sphere
is initially not very well resolved, so that dynamical processes inside
the \hii\ region can lead to an amplification of grid
effects. Resolution studies in two dimensions that we have performed
show that this effect diminishes with increasing resolution.

\subsubsection{Protostellar model}

To trace the collapse of individual fragments, we use a sink particle
technique that we developed for the FLASH code~ \citep{federrathetal09}
(see also \citet{bateetal95}). Here, sink particles are
created if the local density exceeds a critical value
$\rho_\mathrm{crit}$ and the region within the sink particle radius
$r_\mathrm{sink}$ is gravitationally bound and collapsing. For the
high resolution simulations (Run A and Run B), we use
$\rho_\mathrm{crit} = 7 \times 10^{-16}\,$g\,cm$^{-3}$ and
$r_\mathrm{sink} = 590\,$AU, for the lower resolution simulations (Run
Ca and Run Cb), we take $\rho_\mathrm{crit} = 1 \times
10^{-16}\,$g\,cm$^{-3}$ and $r_\mathrm{sink} = 1319\,$AU.  The sink
particles gain mass through accretion of overdense gas
within the accretion radius that is gravitationally bound to it.
The particles interact gravitationally with the gas and with themselves
and are free to move within the simulation box, independent of the
underlying grid structure, i.e. they are treated fully Lagrangian.

We use the properties of these Lagrangian sink particles as sources
for the radiation module. A prestellar model is used to determine
luminosity and effective temperature of the source as function of
protostellar mass and accretion rate. We set the stellar luminosity by
a zero-age main sequence \citep{paxton04} and the accretion luminosity
by interpolation from a more detailed model \citep{hosoomu08}.

The photoionization rate and the photoionization heating rate are set
by the specific mean intensity along a ray
\begin{equation}
J_\nu(r) = \left(\frac{\rstar}{r}\right)^2 \frac{1}{2 c^2}
\frac{h \nu^3}{\exp(h \nu / \kb \tstar) - 1} \exp[-\tauion(r)] .
\end{equation}
Here, $r$ is the distance from the source, $\rstar$ the radius of the
star, $\tstar$ the effective temperature of the star, $\nu$ the
frequency, $c$ the speed of light, $h$ Planck's constant, $\kb$
Boltzmann's constant and $\tauion$ the optical depth for the ionizing
radiation, which is calculated using the absorption coefficient of
atomic hydrogen \citep{rijk06}. Hence, the strength of the ionizing
radiation is totally determined by $\rstar$ and $\tstar$. We use a
zero-age main sequence (ZAMS) model derived from the freely available stellar
evolution code EZ \citep{paxton04} to calculate these quantities.

To also include heating by non-ionizing radiation, we have extended
the raytracing module to calculate Planck mean opacities following the
\citet{krumkleinmckee07} interpolation of data from
\citet{pollack94}. We use the non-relativistic approximation
for the dust heating term \citep[see e.g.,][]{krumkleinmckee07b}
\begin{equation}
\gd = \kp \rho c u
\end{equation}
with the Planck mean opacity $\kp$, the gas density $\rho$ and the
total radiation energy $u$. In the raytracing approximation, the
heating due to stellar radiation is thus given by
\begin{equation}
\gs(r) = \sigma \left(\frac{\rstar}{r}\right)^2 \kp[T(r)] \rho(r) \exp[-\taupm(r)] \tstar^4
\end{equation}
with the Stefan-Boltzmann constant $\sigma$ and the Planck mean opacity $\taupm$.

In addition to the stellar heating, we also include a model for
accretion heating. To this end, we assume that the potential energy of
the gas which is released during accretion is fully converted into
radiation at the accretion radius $\racc$, so that the accretion
luminosity is given by
\begin{equation}
\lacc = G \frac{M \dot{M}}{\racc}
\end{equation}
with Newton's constant $G$, the protostellar mass $M$ and the
accretion rate $\dot{M}$. Since we do not account for diffuse
radiation, the accretion heating also originates from the sink
particles and leads to another heating term of the form
\begin{equation}
\label{eq:ch2accheat}
\gacc(r) = \sigma \left(\frac{\racc}{r}\right)^2 \kp[T(r)] \rho(r) \exp[-\taupm(r)] \tacc^4 .
\end{equation}
To determine the accretion radius $\racc$ and the effective
temperature $\tacc$ of the accretion luminosity, we interpolate the
results of the prestellar evolution model of \citet{hosoomu08} from
the current accretion rate and protostellar mass.

\subsection{Initial Conditions}

We start our calculations with a molecular cloud with a mass of
$1000\,$M$_{\odot}$ and an initial temperature of $30\,$K. The cloud
has a flat inner region with $0.5\,$pc radius, surrounded by a region
in which the density drops as $r^{-3/2}$. The core density is
$1.27 \times 10^{-20}\,$ g\,cm$^{-3}$. The whole simulation box
has an edge length of $3.89\,$pc.
The core is in solid body rotation
with a ratio of rotational to gravitational energy $\beta = 0.05$.
We then run the simulation allowing sink particles to continue forming for $130\,$kyr after
the formation of the first one. Note that the total simulation time is
$0.75\,$Myr.

These initial conditions represent an extrapolation from lower mass
prestellar cores, which are well described by density profiles similar
to Bonnor-Ebert spheres \citep[see e.g.,][]{pirogov09}, towards higher
masses and larger scales. After a simulation runtime of $500\,$kyr,
the flat inner part has totally disappeared, and a centrally condensed
structure with a peak density of $10^{-19}$~g~cm$^{-3}$ and a radius
of $0.65\,$pc has emerged. This is the stage at which infrared dark
clouds are typically observed \citep[see e.g.,][]{beuthhen09}.

In some runs, we choose to suppress secondary sink formation and
instead use a density-dependent temperature floor to prevent runaway
collapse of dense blobs of gas.  We need to resolve the local Jeans
length with $n \geq 4$ cells \citep{truelove97}. To do so, we introduce
a dynamical temperature floor
\begin{equation}
\tmin = \frac{G \mu \mpr}{\pi \kb} \rho (n \Delta x)^2 ,
\end{equation}
where $\mu$ is the mean molecular weight,
$\mpr$ the proton mass, and $\Delta x$ the cell size.

We consider two full resolution simulations with a cell size at the
highest refinement level of 98~AU (11 refinement levels).  In one, Run A, we artificially suppress
disk fragmentation using the dynamical temperature floor, while in the other, Run B, we
permit it and dynamically form secondary sink particles.  In
addition, we ran two half resolution simulations with maximum
resolution of 196~AU (10 refinement levels) and suppressed secondary sink formation.
One of these runs, Run Ca, starts with identical initial conditions to Runs A and~B,
while Run Cb has an additional $m=2$-perturbation of $10\%$ of the gas
density~\citep{bossbod79}. The high-resolution stellar cluster simulation Run~B
reaches $\sim 7.2\times 10^6$ grid cells by the end of the simulation,
and required 
$2.5 \times 10^5$ CPU hours to complete. An overview of the runs with their different model parameters is
given in Tab.~\ref{tab:colsim}.

\begin{table}
\begin{centering}
\begin{tabular}{c||c|c}
Name & Resolution & Multiple Sinks\\
\hline
\hline
Run A & 98~AU & no \\
Run B & 98~AU & yes \\
Run Ca & 196~AU & no \\
Run Cb & 196~AU & no \\
\end{tabular}
\caption{Overview of collapse simulations.}
\label{tab:colsim}
\end{centering}
\end{table}

\section{Accretion}
\label{accretion}

\begin{figure}
\plottwo{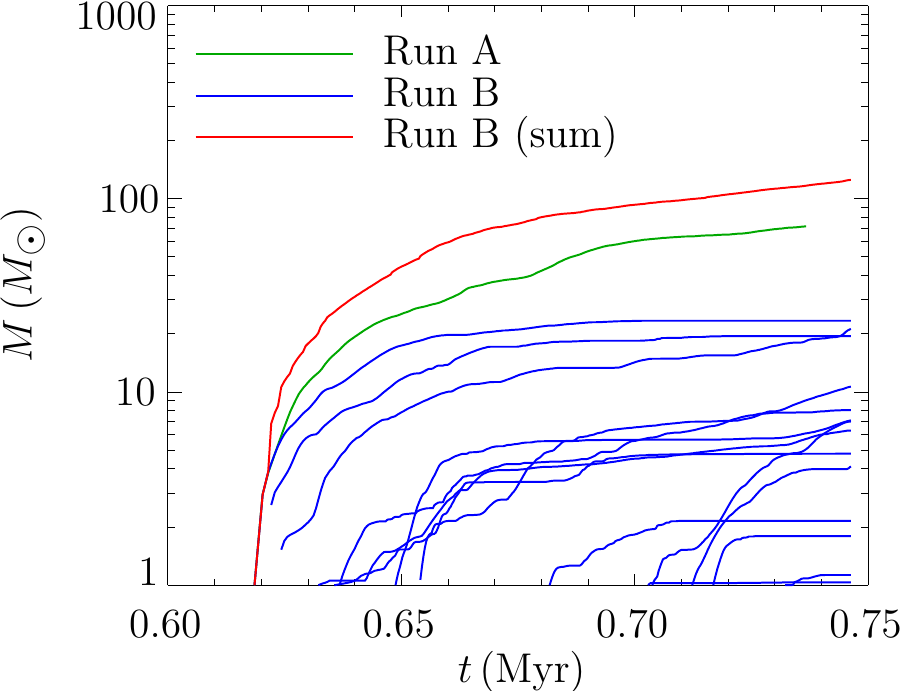}{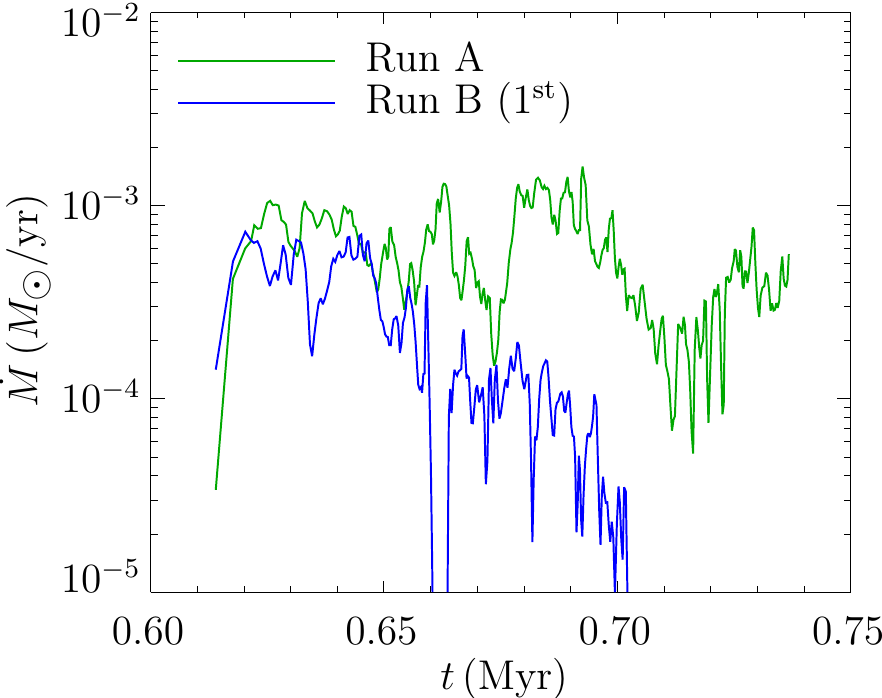}
\label{accretion-rate}
\caption{Accretion history of the single (Run A) and multiple (Run B)
  sink simulations. Run A was stopped at $72\,$M$_{\odot}$, while no
  sink particle in Run B exceeds $25\,$M$_{\odot}$ over the simulation
  runtime. The left hand plot shows the sink particle masses for Run A
  (green), the individual sink masses of Run B (blue) as well as the
  total mass in sink particles in Run B (red). The right hand plot
  shows the accretion rates of the sink particle in Run A (green) and
  sink particle which forms first in Run B (blue), which also turns
  out to end up as the most massive at the end of the simulation.  The
  most massive stars in Run B are those which form early and keep
  accreting at a high accretion rate. While the accretion rate in Run
  A never drops below $10^{-5}\,$M$_{\odot}\,$yr$^{-1}$, accretion
  onto the most massive sink can drop significantly below this value
  and even be stopped totally in Run B.}
\end{figure}

We first examine the accretion histories of our different models.
Figure~\ref{accretion-rate} shows that in Run~A, with only one sink
particle allowed to form, nothing halts accretion
onto the central protostar. It
continues to grow at an average rate of $\dot{M} \approx 5.9
\times 10^{-4}\,$M$_{\odot}\,$yr$^{-1}$ until we stop the calculation
when the star has reached $72\,$M$_{\odot}$.  
The increasingly massive star ionizes the surrounding gas, raising it to high pressure. This
gas breaks out above and below the disk plane, but it cannot halt
mass growth through the disk mid-plane
.

\begin{figure}
\includegraphics[height=170pt]{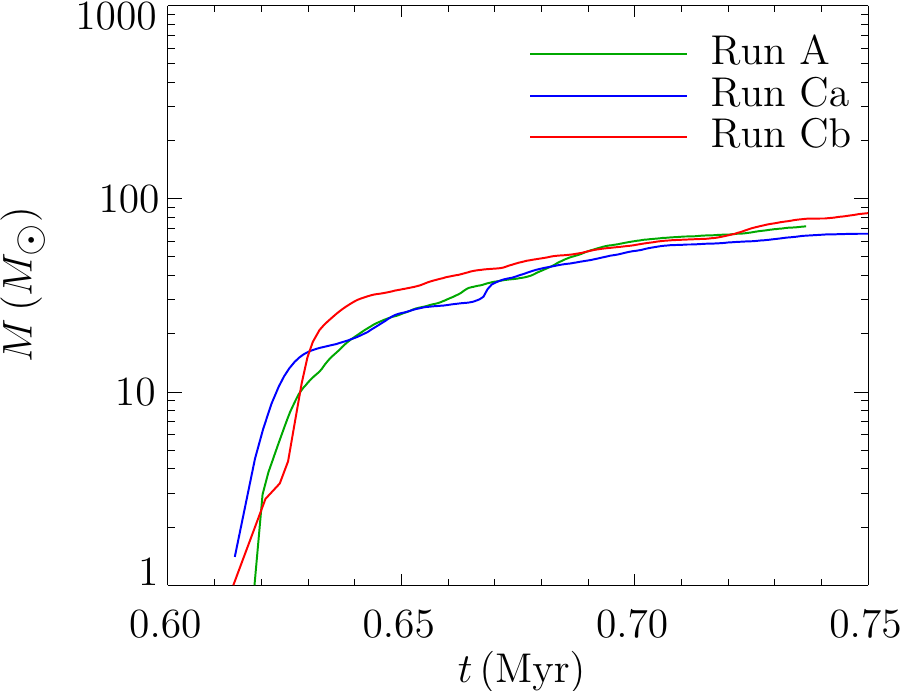}
\includegraphics[height=170pt]{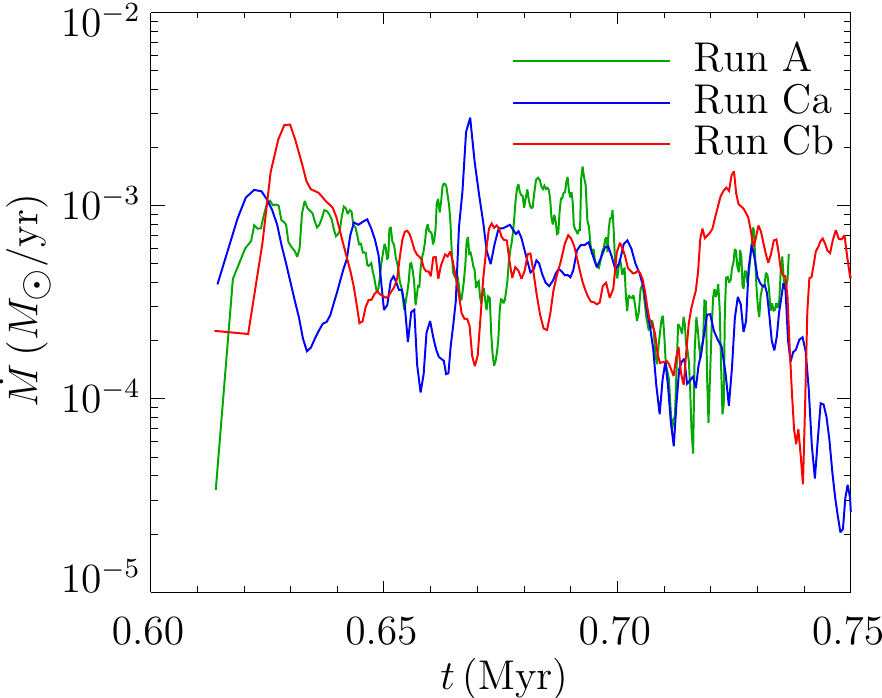}
\caption{ Accretion history of the high resolution (Run A) and low
  resolution simulations with the same initial conditions (Run Ca),
  and with an additional initial perturbation (Run Cb). The accretion
  process proceeds similarly in all three simulations. Run A was
  stopped at $72\,$M$_{\odot}$, Run Ca at $100\,$M$_{\odot}$ and Run
  Cb at $94\,$M$_{\odot}$. The mean accretion rates of $5.9 \times
  10^{-4}\,$M$_{\odot}\,$yr$^{-1}$, $4.6 \times
  10^{-4}\,$M$_{\odot}\,$yr$^{-1}$ and $5.8 \times
  10^{-4}\,$M$_{\odot}\,$yr$^{-1}$, respectively, are also very
  similar.  There is no indication that ionization may stop
  accretion.}
\label{lowsim}
\end{figure}

We ran our lower resolution simulations with a single sink particle,
Runs Ca and Cb, until the sink particles reached masses of
$94\,$M$_{\odot}$ and $100\,$M$_{\odot}$ without any reduction in the
mass accretion rate. Figure~\ref{lowsim} shows the accretion history for
these simulations. One can clearly see that neither the additional
perturbation nor the change in resolution has any significant
influence on the overall accretion behavior. Runs~Ca and~Cb accrete at
mean rates of $4.6 \times 10^{-4}\,$M$_{\odot}\,$yr$^{-1}$ and $5.8
\times 10^{-4}\,$M$_{\odot}\,$yr$^{-1}$, almost identical to the value
for Run A, showing that this quantity is well converged.  These
simulations with a single sink particle strongly suggest that
ionization feedback alone does not limit massive star accretion.

In contrast, in Run B, where multiple sink particles are allowed to form,
two subsequent sink particles form and begin accreting soon after the first one, 
and many more follow within the next $10^{5}\,$yr (see Fig.~\ref{accretion-rate}).
By that time the
first sink has accreted $8\,$M$_{\odot}$. Within another $3 \times
10^{5}\,$yr seven further fragments have formed, with masses ranging
from $0.3\,$M$_{\odot}$ to $4.4\,$M$_{\odot}$ while the first three
sink particles have masses between $10\,$M$_{\odot}$ and
$20\,$M$_{\odot}$, all within a radius of $0.1\,$pc from the most
massive object. Accretion by the secondary sinks terminates the mass
growth of the central massive sinks. Material that moves inwards
through the disk driven by gravitational torques accretes
preferentially onto the sinks at larger radii \citep{bate02}. Eventually,
hardly any gas makes it all the way to the center to fall onto the
most massive objects.
This
fragmentation-induced starvation 
prevents any star from reaching a mass $>25\,$M$_{\odot}$ in this case.

The accretion behavior in Run~B contrasts sharply with competitive accretion models
\citep{bonnell01,bonetal04}, which have no mechanism to turn off accretion
onto the most massive stars. In these models, material falls all the way
down to the massive stars sitting in the center of the gravitational potential
which thereby take away the gas from the surrounding low-mass stars. In our fragmentation-induced
starvation scenario, exactly the opposite happens.

\begin{figure}
\includegraphics[width=400pt]{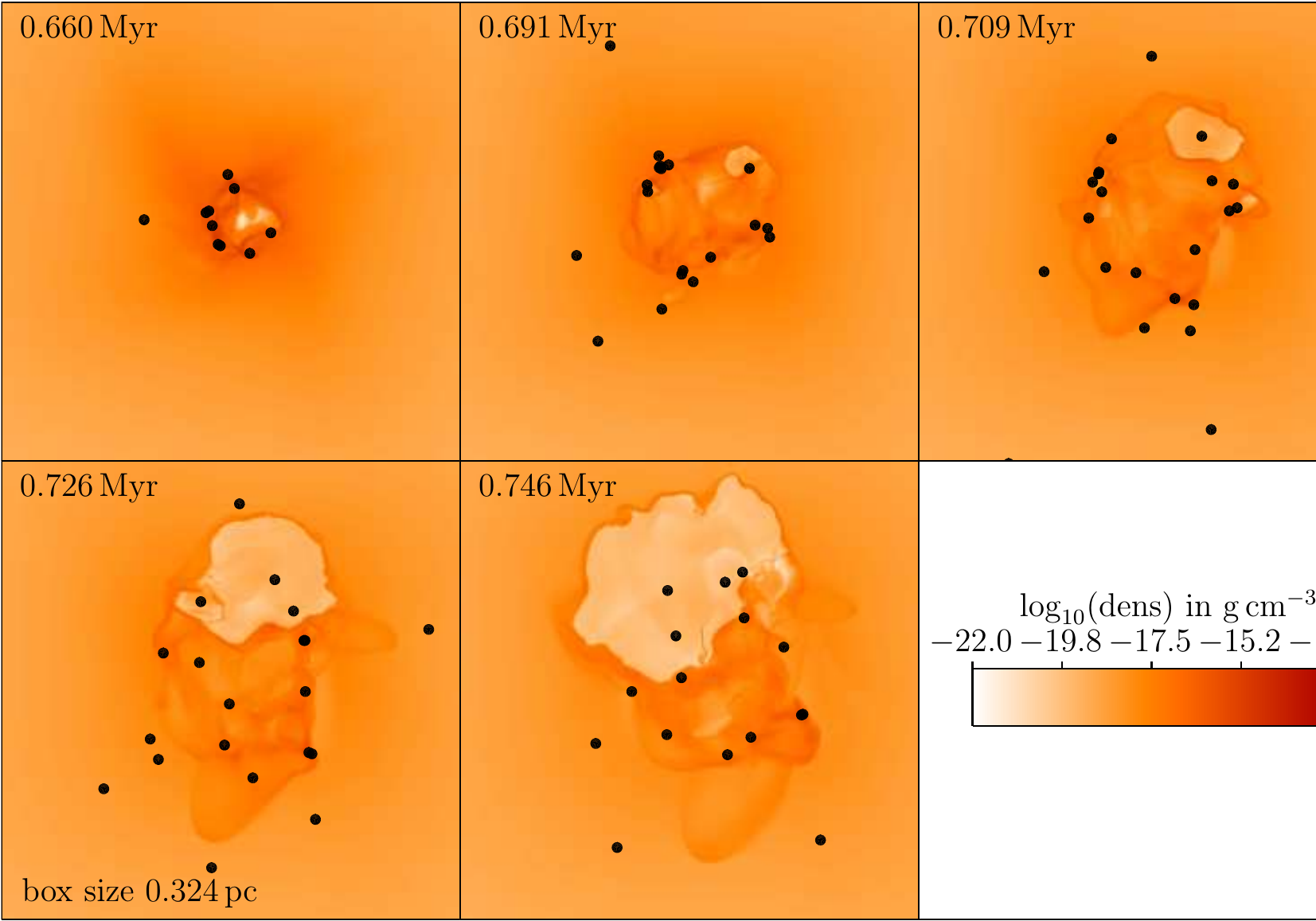}
\caption{ Disk fragmentation in the multiple sink simulation (Run B).
  The figure shows the gas density in slices in the midplane of the
  disk. The filaments in the disk form a successively growing number
  of protostars. As the gas reservoir around the massive protostars is
  exhausted, the thermal pressure of the ionized gas creates a bubble
  around the star. This stops the accretion process. In the course of
  the simulation, the bubbles grow in size and finally merge.  }
\label{diskfragmulti}
\end{figure}

\begin{figure}
\includegraphics[width=400pt]{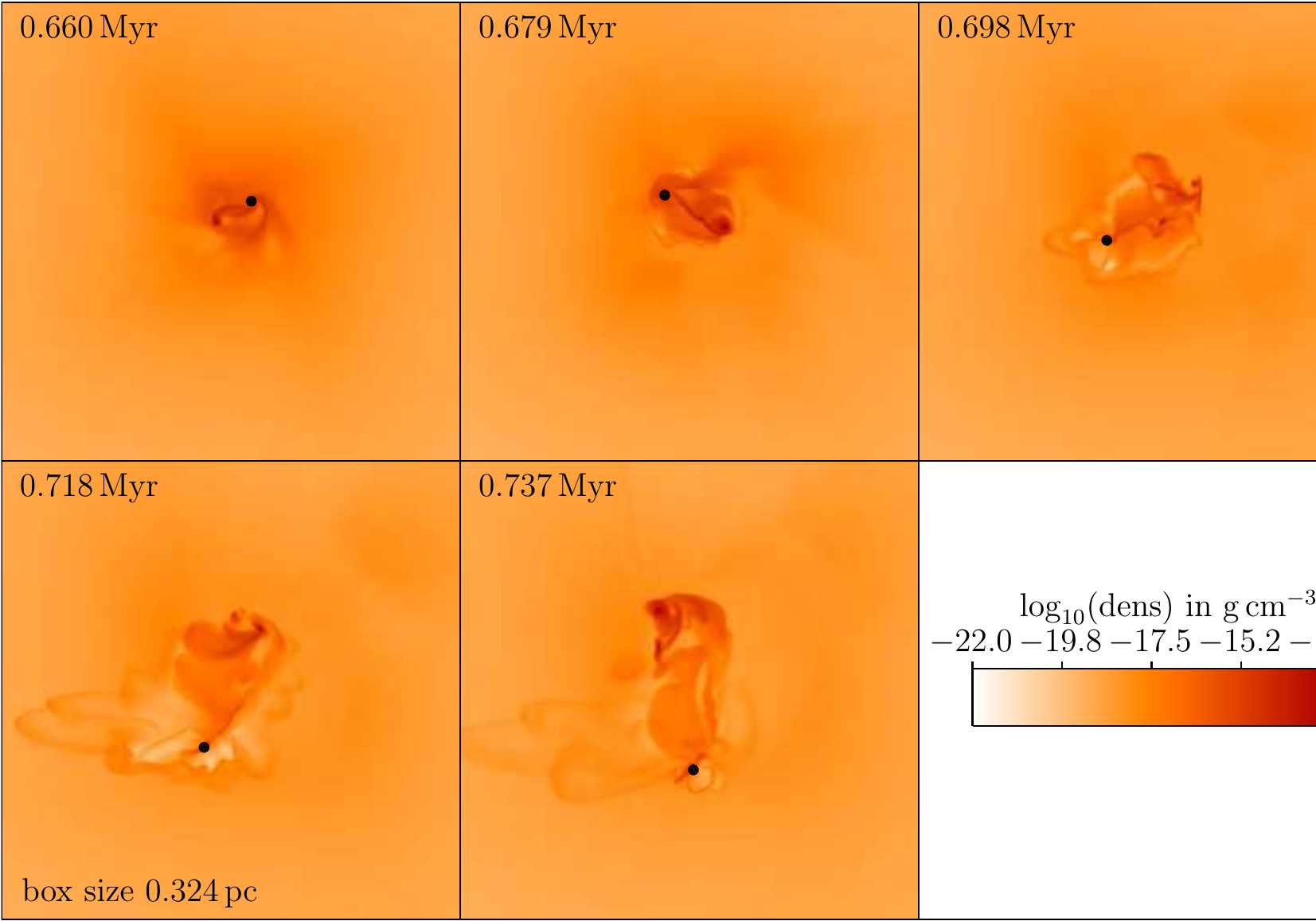}
\caption{ Disk fragmentation in the single sink simulation (Run A).
  The images show the same region as Figure~\ref{diskfragmulti}. This
  time the filaments do not form protostars but blobs of gas which are
  puffed up by the dynamical temperature floor. The ionizing radiation is unable to
  create a bubble in the strong accretion flow large enough to stop
  accretion.  }
\label{diskfragsingle}
\end{figure}

The different behaviour of the multiple sink simulation (Run B) and the
single sink simulation (Run A) can be understood by looking at slices
of density in the disk plane, as shown in Figures~\ref{diskfragmulti}
and~\ref{diskfragsingle}. In Run B, secondary sink particles form in
the disk filaments, which remove gas from the common
reservoir. Consequently, the most massive stars cannot maintain their
high accretion rates. It is this halting of the accretion flow that
allows the ionizing radiation to create a bubble around the massive
protostars.  In contrast, the sink particle in Run A is embedded in an
accretion flow at all times. The ionizing radiation does blow away a
significant fraction of gas from the sink particle, but it is not able
to do so in all directions. There is always a region in the disk
midplane around the sink particle where the gas is dense enough 
that evaporation by the ionizing radiation remains insufficient
to stop the accretion flow.

\section{Bipolar Outflows}
\label{bipolar}

In all runs, the \hii\ regions are 
trapped in the disk plane but drive a bipolar outflow perpendicular to
the disk.  The highly variable rate of accretion onto protostars as
they pass through dense filaments causes fast ionization and
recombination of large parts of the interior of the perpendicular
outflow. The \hii\ regions around the massive protostars do not
uniformly expand, but instead fluctuate rapidly in size, shape and
flux.

\begin{figure}
\includegraphics[width=400pt]{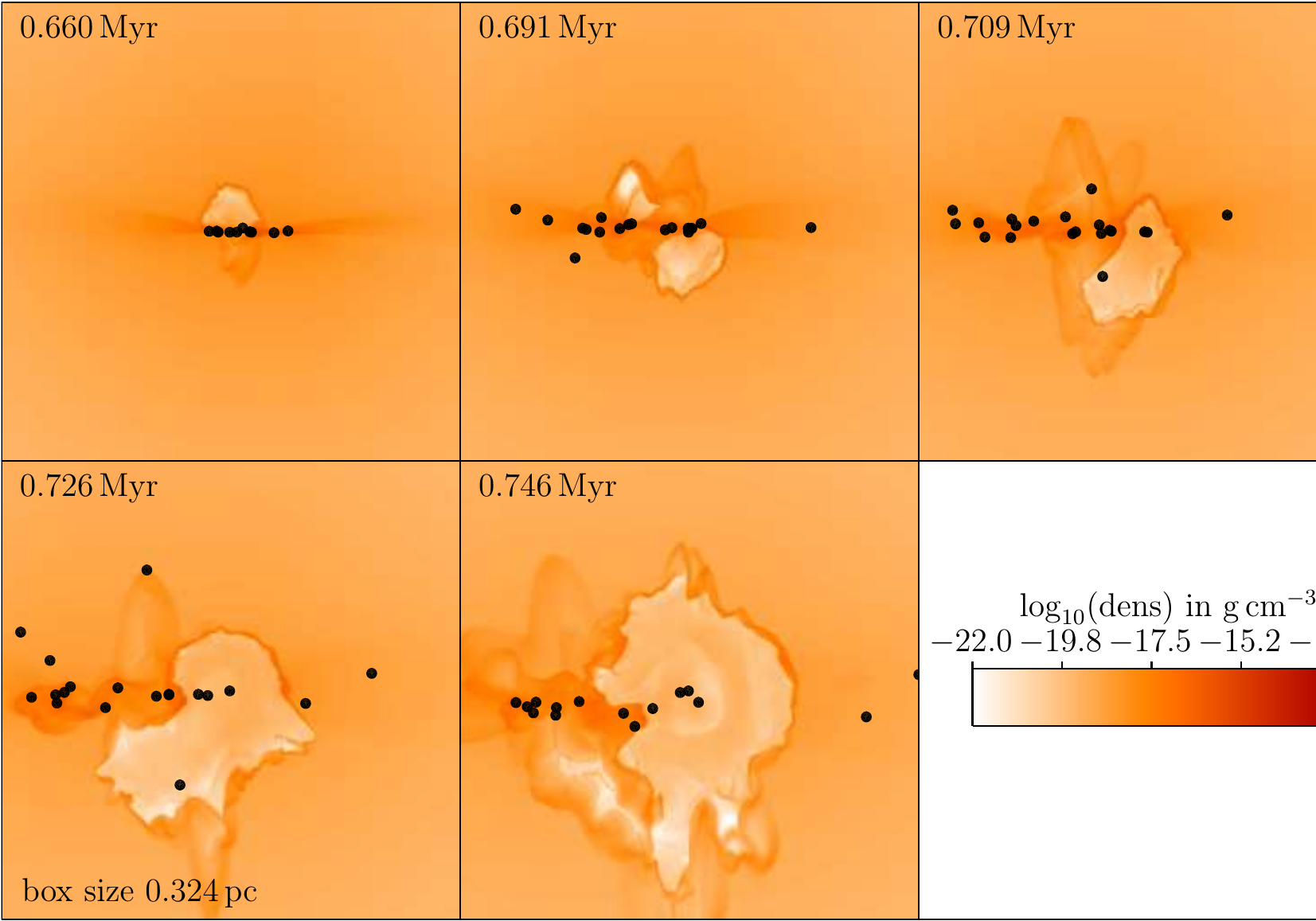}
\caption{Bipolar outflows in the multiple sink simulation (Run B).
The slices show the gas density in a plane through the most massive
sink particle perpendicular to the disk. 
In the beginning of the simulation, the outflow is bipolar but not
symmetric because the filaments prevent 
the thermal pressure to drive simultaneously both lobes of the
outflow. This effect fades when the accretion 
flow drops, and the sink particle can even blow an expanding bubble
although it does not grow in mass anymore in the last 
three frames shown. The scales and times of the images are the same as in
Figure~\ref{diskfragmulti}. The slice plane is chosen such that it includes the sink
particle of interest and is centered around it.}
\label{outflowfragmulti}
\end{figure}

\begin{figure}
\includegraphics[width=400pt]{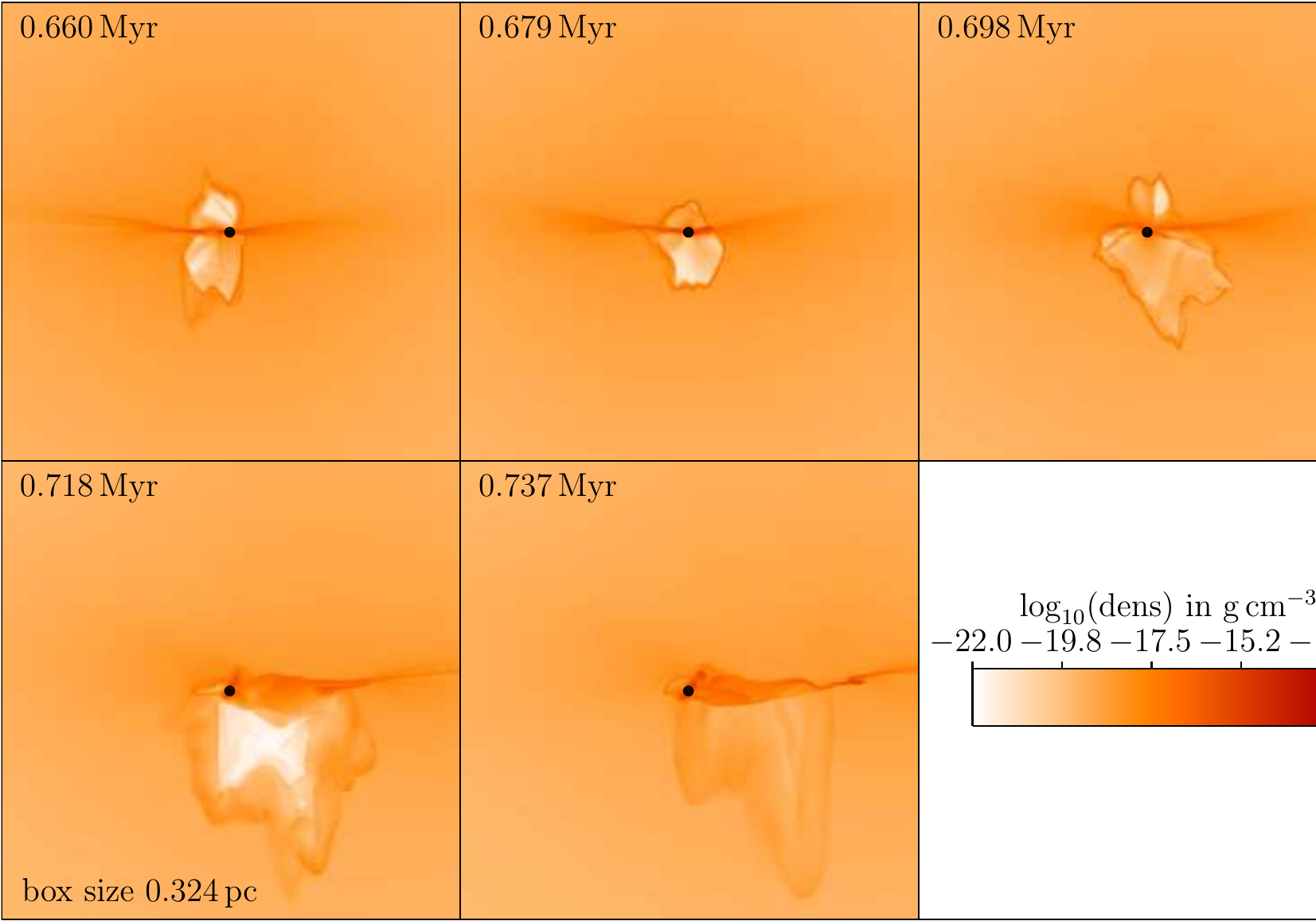}
\caption{ Bipolar outflows in the single sink simulation (Run A). The
  images show that the size of the outflow is not directly related to
  mass of the protostar. Instead, it is given by the time the thermal
  pressure has had to steadily drive it.  If the thermal support is
  lost due to recombination, the outflow gets quenched again. This is
  apparent in the last frame, where the star is the most massive but
  there is no visible cavity around it. The scales and times of the
  images are the same as in Figure~\ref{diskfragsingle}. The slice
  plane is chosen such that it includes the sink particle of interest
  and is centered around it.  }
\label{outflowfragsingle}
\end{figure}

The differences between the single and multiple fragment Runs A and
Run B are also evident from the characteristics of the bipolar
outflows. Figures~\ref{outflowfragmulti} and~\ref{outflowfragsingle}
show density slices perpendicular to the disk plane for both
simulations. Figure~\ref{outflowfragmulti} shows the outflows driven by the most
massive sink particle in Run B. The efficient shielding by the
filaments and the motion of the sink particle hinder the thermal
pressure of the ionized gas from driving a symmetric bipolar outflow at
early times. The shielding of the ionizing radiation leads to a drop
in the thermal pressure which drives and supports the outflow, and
thus the outflow can be quenched again.  At later times, the strong
accretion flow onto the sink particle stops, and this allows the sink
particle to drive a much larger outflow, which in the end has the form
of a bubble.

The situation for the sink particle in Run A is presented in
Figure~\ref{outflowfragsingle}. The general properties of the outflow
are very similar, but the big difference is that here the accretion
flow never stops. On the contrary, Figure~\ref{accretion-rate} shows
that the accretion rate increases towards the end of the
simulation. This strong accretion flow fully absorbs the ionizing
radiation, so the sink particle is unable to build up a bubble, even though
it has more than three times the mass of the most massive
star in Run B.

Observed outflows \citep{arceetal07} from O-type protostars with ages
$> 10^5$ yr tend to be much wider than those from low- or even intermediate-mass
protostars. When resolved, these outflows show a disordered appearance that may be the
outcome of multiple powering sources (including \hii\ regions) and
non-steady driving. The pressure-driven outflows produced by \hii\
regions in our
simulations reproduce these broad, disordered outflows well. Collimated outflows
from less massive protostars without \hii\ regions are likely magnetically
driven, though \citep[see e.g.,][]{banerjee06b}. A more detailed model
linking these two types of outflows is out of the scope of this paper.

\section{Comparison to Observations}
\label{comparison}
We can directly compare our models with radio observations of
free-free continuum, hydrogen recombination lines, and NH$_3(3,3)$
inversion lines by generating synthetic maps.

\subsection{Methods}
We generate radio continuum maps from our numerical data by
integrating the radiative transfer equation in the Rayleigh-Jeans
limit while neglecting scattering \citep{kraus66,gorsor02}. To this end,
we calculate for every cell in the domain the free-free absorption
coefficient for atomic hydrogen
\begin{equation}
\alpha_\nu = 0.212\left(\frac{\nel}{1\,\mathrm{cm^{-3}}}\right)^2 
\left(\frac{T_\mathrm{e}}{1\,\mathrm{K}}\right)^{-1.35} \left(\frac{\nu}{1\,\mathrm{Hz}}\right)^{-2.1} \mathrm{cm^{-1}},
\end{equation}
where $\nel$ is the number density of free electrons, $T$ the gas
temperature and $\nu$ the frequency. The optical depth 
at distance $r$ is then
\begin{equation}
\tau_\nu = \int_0^r \alpha_\nu\,\mathrm{d}s .
\end{equation}
Without scattering, the radiative transfer equation for the brightness
temperature can be directly integrated and yields 
\begin{equation}
T(\tau_\nu) = \mathrm{e}^{-\tau_\nu} \int_0^{\tau_\nu} \mathrm{e}^{\tau'_\nu} T(\tau'_\nu)\,\mathrm{d} \tau'_\nu  .
\end{equation}
If $\Omega_\mathrm{S}$ is the solid angle subtended by the beam, the resulting flux density is
\begin{equation}
F_\lambda = \frac{2 \kb\, T}{\lambda^2} \Omega_\mathrm{S} .
\end{equation}
Following the algorithm described in 
\citet{maclowetal91}, we convolve the resulting image with the
beam width and add some noise according to the telescope parameters.

For a time-series comparison at high temporal resolution, we select a
few time intervals in the multi-sink run (Run B) during which the
continuum emission from a particular, well-isolated, \hii\ region
varies strongly. We then produced 2~cm continuum maps every $\sim 10$ yr
from the simulation output and measured the properties of the region, including
the observational scale length $H$, the flux $F_{2\mathrm{cm}}$, and
the accretion rate of the protostar $\dot{M}_\mathrm{sink}$ that
powers the \hii\ region.  $H$ is defined as the equivalent diameter of
a circle with the same area as the emission. $F_{2\mathrm{cm}}$ is
obtained by integrating the intensity of the maps over solid
angle. $\dot{M}_\mathrm{sink}$ is the accretion rate of the sink
particle. 

We generate synthetic observations of line emission using our
radiative transfer code MOLLIE \citep{keto90}. This code generally uses
the Accelerated Lambda iteration algorithm \citep{rybhum91} although in
this investigation we compute the level populations of NH$_3$ assuming
local thermodynamic equilibrium (LTE) and assume optically thin
conditions for the H53$\alpha$ recombination line. The computation is
done on 4 nested grids with spatial scales ranging from 80 to 640 AU
and covering 0.2 pc. The maps of the NH$_3$ and H53$\alpha$ velocities
show the first moment, in the case of NH$_3$ integrating over the main
hyperfine line of the (3,3) transition.

\subsection{Results}
\begin{figure}
\epsscale{.80}
\plotone{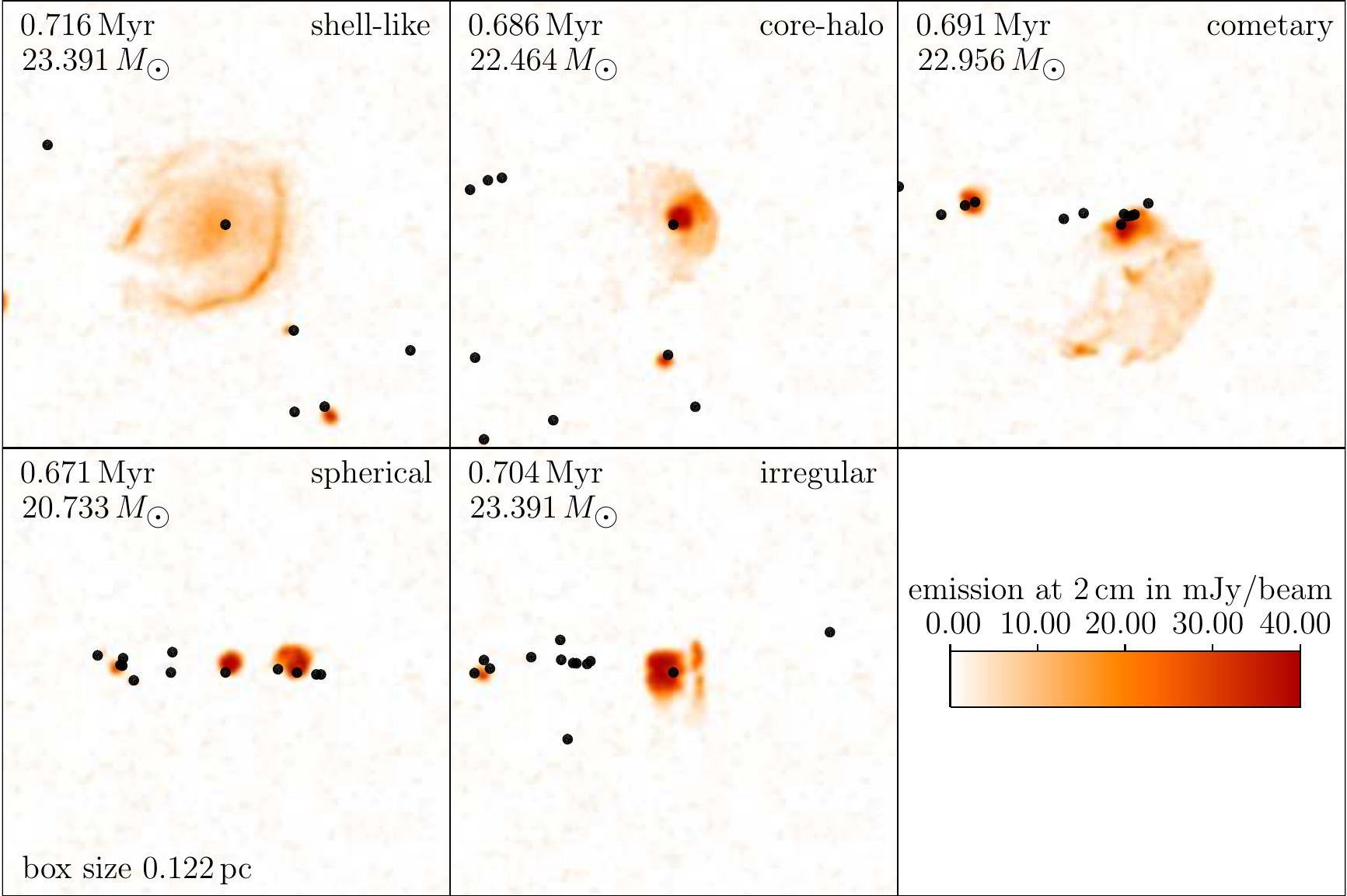}
\caption{\hii\ region morphologies. This figure shows ultracompact \hii\ regions around massive protostars in Run B at different
time steps and from different viewpoints. The cluster is assumed to be $2.65\,$kpc away, the full width at half maximum of the beam is
$0\farcs14$ and the noise level is $10^{-3}$Jy. This corresponds to typical VLA parameters at a wavelength of $2\,$cm.
The protostellar mass of the central star which powers the \hii\ region is
given in the images. The \hii\ region morphology is highly variable in time and shape, taking the
form of any observed type \citep{woodchurch89,kurtzetal94} during the
cluster evolution.
\label{radio-morphology}
}
\end{figure}

Our simulated observations of radio continuum emission reproduce the 
morphologies reported in surveys of ultracompact \hii\
regions \citep{woodchurch89,kurtzetal94}. Figure~\ref{radio-morphology} shows
typical images from Run~B, and the Online Material contains a
movie of the entire run. The \hii\ regions in our model fluctuate
rapidly between different shapes while accretion onto the protostar
continues. When the gas reservoir around the two most massive stars is
exhausted, their \hii\ regions merge into a compact \hii\ region
similar to the extended envelopes often observed around ultracompact \hii\
regions \citep{kimkoo01,kurtzetal99}. Our results suggest that the lifetime problem
of ultracompact \hii\ regions \citep{woodchurch89} is only apparent.
Since \hii\ regions embedded in accretion flows are continuously
fed, and since they flicker with variations in the flow rate,
their size does not depend on their age until late in their lifetimes.

The same mechanisms that cause the different accretion behavior (\S~\ref{accretion})
and bipolar outflows (\S~\ref{bipolar}) for Run~A and Run~B also lead
to differences in the \hii\ region morphologies. One the one hand,
the permanent accretion flow in Run~A quenches the \hii\ region in the disk
plane. On the other hand, the higher stellar mass in Run~A leads to larger \hii\ regions
perpendicular to the disk plane. Since the second effect dominates, the bottom line is
that the \hii\ regions in Run~A are generally larger than in Run~B.
Hence, multiple sink formation is not only necessary to model the gas fragmentation
correctly, it is also crucial to reproduce the large relative fraction
of spherical \hii\ regions observed in surveys \citep{woodchurch89,kurtzetal94}.
A detailed investigation of the \hii\ region morpholgies, including a comprehensive
statistical analysis for both simulations, will be presented in a follow-up paper.

\begin{figure}
\plottwo{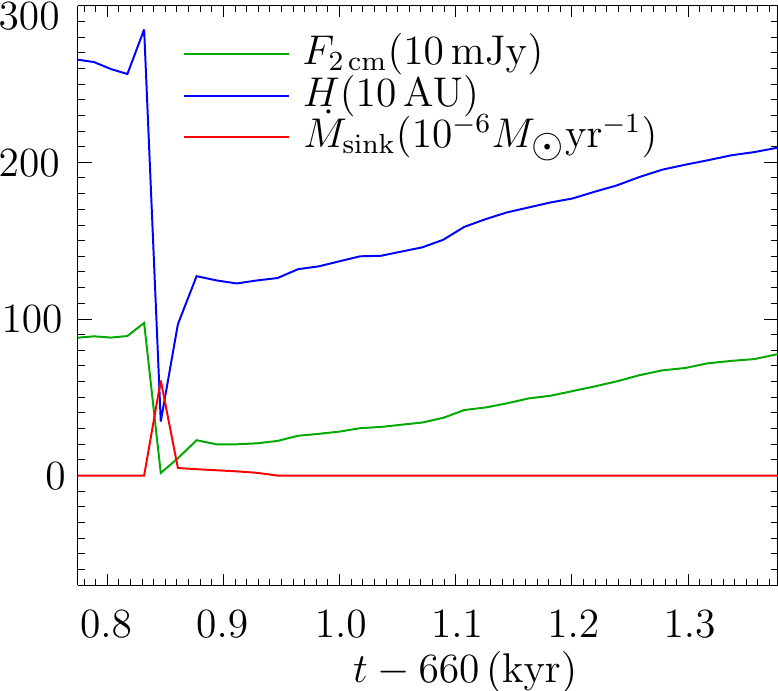}{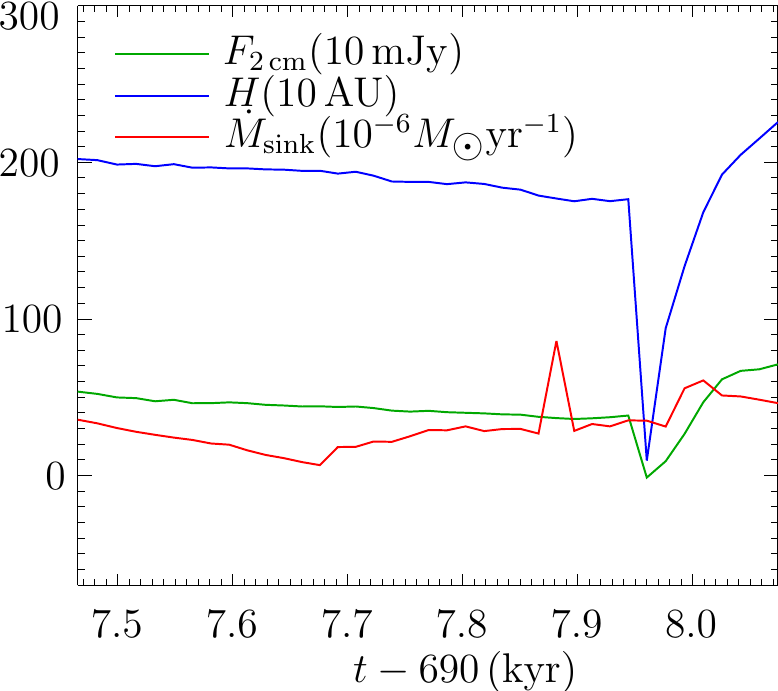}
\caption{Large amounts of molecular gas accreting onto an \hii\ region can cause a sudden
decrease in its size and flux. This figure shows the 2~cm continuum flux $F_{2\mathrm{cm}}$
(in units of 10 mJy), the characteristic size of the \hii\ region $H$ (in units of 10 AU),
and the rate of accretion to the star $\dot{M}_\mathrm{sink}$ (sink particle, in units
of $10^{-6} M_\odot$ yr$^{-1}$) for two accretion events at high temporal resolution.
\label{timeseries}}
\end{figure}

To compare more directly to observations of the time variability of
\hii\ regions, we analyzed a few time intervals of interest at a time
resolution of $\sim 10$ yr. We find that when the accretion rate to
the star powering the \hii\ region has a large, sudden increase, the
ionized region shrinks, and then slowly re-expands.  This agrees with
the contraction, changes in shape, or anisotropic expansion observed
in radio continuum observations of ultracompact \hii\ regions over
intervals of $\sim$ 10$\,$yr
\citep{francheretal04,rodrigetal07,galvmadetal08}. Figure~\ref{timeseries}
shows that the sudden accretion of large amounts of material is
accompanied by a fast decrease in the observed size and flux of the
\hii\ region.
In the left panel, the \hii\ region is initially relatively
large, and accretion is almost shut off. A large (from 0 to $6 \times 10^{-5} M_\odot$ yr$^{-1}$),
sudden accretion event causes the \hii\ region to shrink and decrease in flux. The star at
this moment has a mass of $19.8\,$M$_\odot$. In the right panel, the star has a larger mass
($23.3\,$M$_\odot$), the \hii\ region is initially smaller, and the star is constantly
accreting gas. The ionizing-photon flux appears to be able to ionize the infalling gas stably,
until a peak in the accretion rate by a factor of three and the
subsequent continuous accretion of gas 
makes the \hii\ region shrink and decrease in flux. The \hii\ region does not
shrink immediately after the accretion peak because the increase is relatively mild and the
geometry of the infalling gas permitted ionizing photons to escape in one direction. Our
results show that observations of large, fast changes in ultracompact \hii\
regions \citep{francheretal04,rodrigetal07,galvmadetal08} are
controlled by the accretion process. 
Both the scale length and flux decrease at rates of 5--7
\% per year in our model,
agreeing well with observed fluctuations of 2--9 \% per year
\citep{francheretal04,galvmadetal08}. 
Shortly after the minimum values are reached, the \hii\ regions
re-expand, on timescales $\sim 10^2$~yr.

\begin{figure}
\epsscale{.80}
\plotone{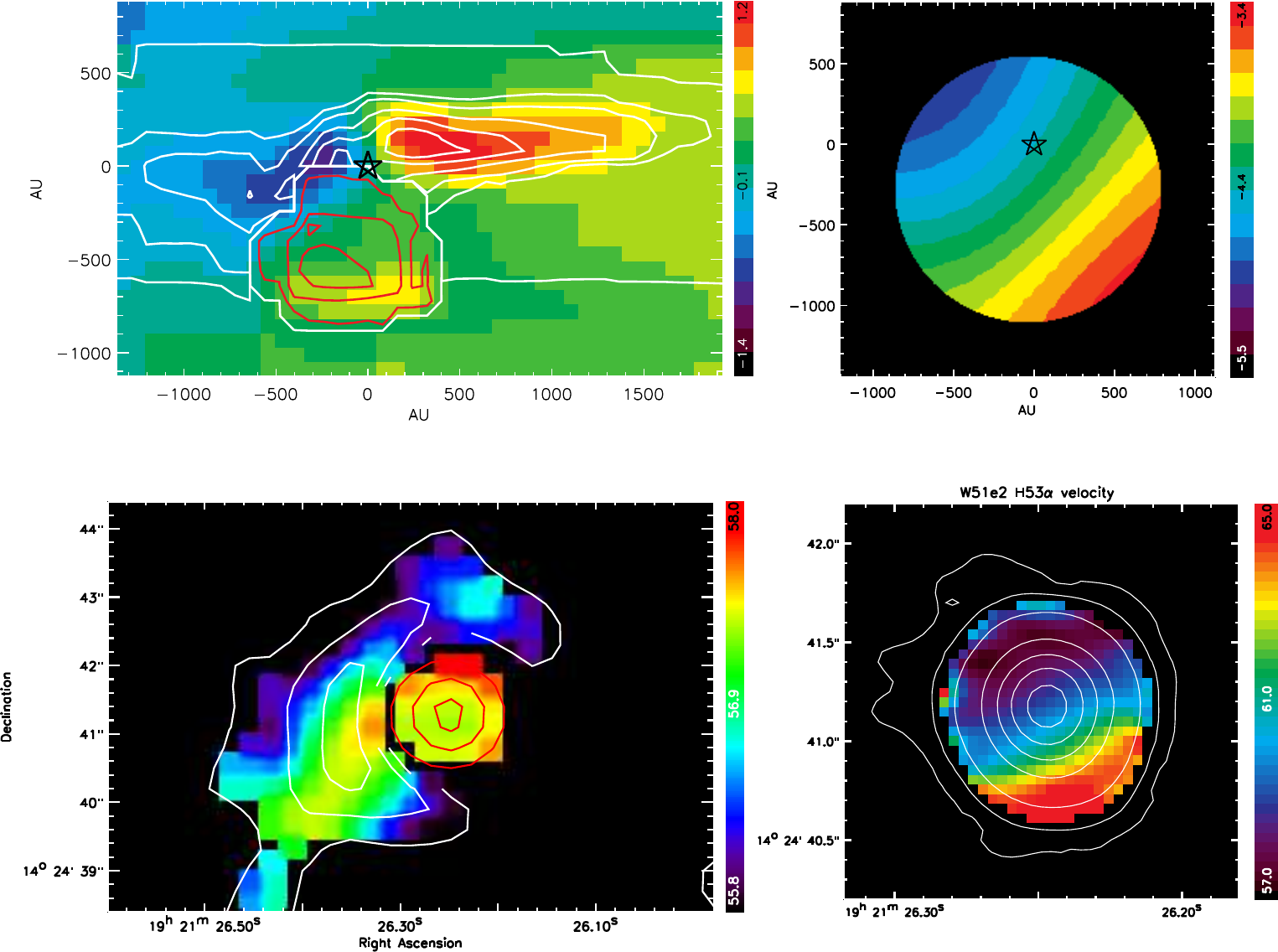}
\caption{Comparison of line and continuum emission simulated from the model (upper panels) and actually observed
from the W51e2 region (lower panels). The left panels show the
NH$_3$(3,3) line emission strength in white contours, the
molecular line velocities as the background color, and the 1.3 cm free-free continuum from ionized
gas in red contours.  The right panels show the H53$\alpha$ recombination line velocities from the ionized gas.
The simulated H53$\alpha$ observations (upper-right) are convolved to a spatial resolution of 750 AU (FWHM) to
better match the relative resolution of the actual observations (lower-right).
The molecular line observations are from
\citet{zhangetal98}. The H53$\alpha$ observations are from \citet{ketoklaas08}.
\label{w51}
}
\end{figure}

We can compare our models to a well-studied
example of an ultracompact \hii\ region,
W51e2 \citep{zhangetal98,ketoklaas08}.  Although we did not choose
parameters specifically to model this region, the
comparison is nevertheless revealing.  In Figure~\ref{w51} we show
simulated and observed maps of NH$_3(3,3)$ emission, 1.3~cm thermal
continuum emission, and the H53$\alpha$ radio recombination line. The
simulated maps were made at a time when a 20 M$_\odot$ star has formed
in Run B.

In the simulated observations, the origin of
the coordinates is referenced to the location of the 20 M$_\odot$ protostar at the center of the accretion flow as marked.
The accretion flow and disk are viewed edge-on. The spatial scale in the observations (lower panels) is 5100 AU per
arc sec assuming W51e2 is at a distance of 5.1~kpc \citep{xuetal09}. The color bar on the right of each figure shows the molecular
velocities in km s$^{-1}$. The velocities of the observations include the LSR velocity of W51e2, approximately 57 kms$^{-1}$.
The white contour levels  are 50\% through 90\% of the peak brightness temperature of 71 K (upper left) and
0.1, 0.2 and 0.3 Jy beam$^{-1}$ kms$^{-1}$ (lower left). The red contour levels are 30\%, 70\%, and 95\% of the peak
1.3 cm continuum brightness temperature of 8902 K (upper left) and 0.07, 0.14, and 0.22 Jy/beam (lower left).
The white contours on the H53$\alpha$ observations (lower-right) show the 7 mm continuum emission
at 2, 4, 10, 30, 50, 70, and 90\% of the peak emission of 0.15 Jy beam$^{-1}$. The molecular line observations are from
\citet{zhangetal98}. The H53$\alpha$ observations are from \citet{ketoklaas08}.
Both observations have a beamsize of about
1$^{\prime\prime}$. Coordinates are in the B1950 epoch.

The brightest NH$_3(3,3)$ emission indicates the dense accretion disk
surrounding the most massive star in the model, one of several within
the larger-scale rotationally flattened flow.  The disk shows the
signature of rotation, a gradient from redshifted to blueshifted
velocities across the star. A rotating accretion flow is identifiable in the
observations, oriented from the SE (red velocities) to the NW (blue
velocities) at a projection angle of 135$^\circ$ east of north
(counterclockwise).

The 1.3 cm radio continuum traces the ionized gas, which in the model
expands 
downward perpendicularly to
the accretion disk, down the steepest
density gradient. As a result, the simulated map shows the brightest
radio continuum emission just off the mid-plane of the accretion disk,
separated from the central star, rather than surrounding it
spherically. In the observations of W51e2, continuum emission is indeed offset
from the accretion disk traced in ammonia. The NH$_3$(3,3) in front of
the \hii\ region is seen in absorption and red-shifted by its inward
flow toward the protostar. The density gradient in the ionized flow
determines the apparent size of the \hii\ region. Therefore the
accretion time scale determines the age of the \hii\ region rather
than the much shorter sound-crossing time of the \hii\ region.

Photoevaporation of the rotationally flattened molecular accretion flow supplies the ionized
outflow. Therefore, the ionized gas rotates as it flows outward,
tracing a spiral. An observation that only partially resolves the spatial
structure of the ionized flow sees a velocity gradient oriented in a
direction between that of rotation and of outflow, as shown in the
simulated observation. The observed H53$\alpha$ recombination
line \citep{ketoklaas08} in Figure~\ref{w51} indeed shows a velocity
gradient oriented between the directions of rotation and the outflow.

\section{Discussion}
\label{discussion}

\subsection{Context}
In this work, we focus on the impact of feedback from radiative
heating by both ionizing and non-ionizing radiation during massive
star formation. Our work is complementary to recent simulations that
neglect ionizing radiation but include other effects.

For example, we use a ray-tracing algorithm to describe ionizing and non-ionizing radiation while
\citet{yorke02, krumkleinmckee07,krumetal09,pricebate09} rely on flux-limited diffusion 
to follow the radiation field.
Like us, these models include heating from accretion luminosity.  
They neglect heating by ionizing radiation, but do
include radiation pressure and scattering effects.  Radiation pressure
in particular becomes important at the
10~AU scale.  These models therefore simulate rather
smaller scales than we do, starting from a $\sim100\,M_\odot$ core of only
$\sim 0.1\,$pc radius, as opposed to our more massive clump with
$1000\,M_\odot$ and a $1.6\,$pc radius.  As a result, our calculations
are better geared towards describing the cluster-forming scale as 
opposed to modeling in detail the mass growth of individual protostars through 
their inner accretion disks. 
\citet{yorke02,krumkleinmckee07,krumetal09,sigalottietal09} show that radiation pressure cannot stop
accretion onto the massive protostar and is dynamically unimportant
except for massive star clusters near the Galactic center or in
starburst galaxies~\citep{krummatzner09}. \citet{krumkleinmckee07}
initialized a microturbulent flow within their dense core, but
\citet{krumetal09} found that it made little difference by the time
material had collapsed to the size of the disk. 
In addition, \citet{pricebate09} 
include magnetic fields and find that they provide some support for  low-density
gas, thus reducing the amount of matter available for accretion. 
\citet{banerjee06b}, \citet{machidaetal05b}, \citet{hennteys08}, \citet{hennfrom08} and \citet{hennciar09} furthermore show that
the presence of magnetic fields 
can dramatically alter the shape and size of accretion
disks and strongly suppress the formation of close binary systems.
These last results, may, however, be an artifact of the neglect of
turbulent and ambipolar diffusion within the disks in these models.
\citet{dalebonnelle08} and \citet{wangetal09}, on the other hand,
focused on mechanical feedback through winds and bipolar outflows,
without and with magnetic fields respectively,
but neglect radiative heating entirely.  They find that outflows
further limit accretion, but do not fundamentally change the picture
of cluster formation in a gravitationally unstable accretion
flow. Ultimately, the quantitative characterization of the star formation
efficiency will require treating the effects of radiation, magnetic fields, and
outflows simultaneously.  

\subsection{Limitations}

As noted above, we have neglected a number of physical processes,
including radiation pressure, scattering, magnetic fields, winds and
bipolar outflows, and turbulent initial conditions. In addition, we
here discuss other approximations that we have made.

Our prestellar model is designed such that the number of UV photons
emitted by the protostar is overestimated for three reasons. First, we
use a ZAMS model to set the ionizing luminosity, while the models of
\citet{hosoomu08} suggest that the protostellar radii of massive stars
can be larger than the ZAMS value, leading to a smaller
ionizing luminosity compared to the ZAMS value.  Second, we neglect
the aborption of UV photons by dust inside the \hii\ region. Thus, the
sizes of our \hii\ regions represent upper limits to the real
values. Third, the dense inner disk around each protostar is subsumed
in the sink particle, allowing radiation to propagate along the disk
midplane to the edge of the sink particle, where densities will be
lower.

However, as discussed in \S~\ref{accretion}, we find that, although we
overestimate the strength of the ionizing radiation, it is unable to
stop accretion onto a single massive star. Similarly, in
\S~\ref{comparison} we showed that the strength of the accretion flow
restricts the expansion of the \hii\ region, which solves the lifetime
problem of ultracompact \hii\ regions \citep{woodchurch89}. Both
results are robust in the sense that the impact of the ionizing
radiation on the accretion flow and the sizes of the \hii\ regions
would even be smaller if we used a more detailed modeling of the
protostellar evolution and environment, and included dust absorption.

We also note that, although we nominally use a simplified treatment of
radiative losses in the optically thick regime, this will not have an
impact on the dynamics of the simulated system, as we expect that the
gas will become optically thick at densities of $\sim
10^{-13}$\,g\,cm$^{-3}$~\citep[e.g.][]{larson69}, which is two orders
of magnitude above the threshold density for sink creation.

\section{Conclusions}
\label{conclusions}

Our results suggest that the accretion flow onto massive stars cannot
be stopped by ionizing radiation (\S~\ref{accretion}).  Instead, the growth of massive
stars is halted by competition from lower-mass companions that form
by gravitational fragmentation in the dense, rotationally flattened,
accretion flow.  This process, which we call fragmentation-induced starvation, limits
the maximum stellar mass of the highest-mass star in our simulations.

Our models lead to a self-consistent picture of massive star formation
where massive stars form through (unsteady) mass growth within a
gravitationally unstable accretion flow. Radiation feedback from the
accretion luminosity and the protostellar surface heats up the
environment in which the first massive protostar forms
\citep{krumkleinmckee07}. This increases the local Jeans mass, so
that subsequent fragments in the neigborhood of this star are
relatively massive, naturally resulting in a cluster of massive
stars. 

Unlike in the competitive accretion model~\citep{bonnelletal06}
where the most massive protostar continues to accrete from the common
gas reservoir, in our model accretion is shut off by the subsequent
fragmentation of the gas because the lower-mass companions intercept
material that would otherwise accrete onto the central star, starving
it of material. Fragmentation-induced starvation will only occur in
regions with high enough density for many massive fragments to form
despite accretion heating.
In our models, the most massive star starts to form early
and completes accretion while the whole cluster is still growing, contrary
to standard competitive accretion models in which the most massive
star grows over the whole cluster evolution timescale.
On the other hand, competition for a common reservoir continues to define the
top end of the initial mass function, unlike in models that rely on
the clump mass spectrum to define the stellar initial mass function.

Our models show (\S~\ref{comparison}) that the \hii\ regions formed
around massive young stars flicker but do not grow systematically
while heavy accretion from the protostellar core continues. This
behavior seems able to explain the many puzzles raised by observations of
ultracompact \hii\ regions. The chaotic interaction between
gravitationally unstable gas filaments in the accretion disk and
ionizing radiation from the young stars creates a great variety of
different ultracompact \hii\ region morphologies. We find shell-like,
core-halo, cometary, spherical and irregular morphologies in a single
simulation, depending on the current flow field and viewing angle. Our
models show behavior consistent with observations of shrinking \hii\
regions with changes in size or flux of a few percent per year, as a
consequence of denser material being accreted the star and shielding
the ionizing radiation.  Shielding by the disk also produces
uncollimated bipolar outflows in our models (see \S~\ref{bipolar}).  These
results lead to an evolution scenario for \hii\ regions that is
substantially different from the classical picture of monotonically
expanding hot bubbles. The surprising dynamics of \hii\ region
flickering resolves the lifetime problem for ultracompact \hii\
regions, since the size of the \hii\ region remains unrelated to the
age of the protostar, as long as fast accretion continues. A detailed
comparison of simulated observations of our model with the
ultracompact \hii\ region W51e2 demonstrates that our model reproduces
distinctive features of massive star forming regions such as \hii\
regions expanding asymmetrically away from the protostar down the steepest
density gradient, and outward-twisting, spiraling, ionized flows. All
these observations can be understood as consequences of ionization
within a gravitationally unstable accretion flow.

\acknowledgments

       T.P. is a fellow of the
       International Max Planck Research School for Astronomy and Cosmic
       Physics at the University of Heidelberg and the Heidelberg Graduate
       School of Fundamental Physics. He also acknowledges support from an Annette Kade Fellowship for his
       visit to the American Museum of Natural History.
       We also thank the {\em Deutsche
       Forschungsgemeinschaft} (DFG) for support via the Emmy Noether Grant BA
       3607/1, via the priority program SFB 439 ``Galaxies in the Early Universe''
       as well as via grants KL1358/1, KL1358/4 and KL1358/5. In addition, we
       also acknowledge partial support from a Frontier grant of Heidelberg University
       funded by the German Excellence Initiative, from the German {\em
       Bundesministerium f\"{u}r Bildung und Forschung} via the ASTRONET project
       STAR FORMAT (grant 05A09VHA) and  from the {\em Landesstiftung
       Baden-W{\"u}rttemberg} via their program International
       Collaboration II.  M.-M.M.L. thanks the Max-Planck-Institut f\"ur
       Astronomie and the Institut f\"ur Theoretische Astrophysik der
       Universit\"at Heidelberg for hospitality, and the U.S. National
       Science Foundation for funding under grant AST08-35734.
       We acknowledge computing time at the Leibniz-Rechenzentrum in Garching
       (Germany), the NSF-supported Texas Advanced Computing Center (USA), and
       at J\"ulich Supercomputing Centre (Germany). The FLASH code was in part
       developed by the DOE-supported Alliances Center for Astrophysical Thermonuclear
       Flashes (ASCI) at the University of Chicago. We thank C. Federrath for important
       improvements to the sink particle module.

\end{document}